\documentclass[12pt]{article}  
\usepackage{times} 
\usepackage{fleqn} 
\usepackage{amssymb,bm} 
\usepackage{epsfig,graphicx,color}  
\usepackage{import} 

\begin{document}

\addtolength{\baselineskip}{0.20\baselineskip}

\vspace{48pt}


\centerline{\Huge Monte Carlo simulations of the NJL model near}
\bigskip
\centerline{\Huge the nonzero temperature phase transition}
\vspace{28pt}

\centerline{\bf Costas Strouthos$^a$ and
Stavros Christofi$^b$}
                                                                                                                                      
\vspace{15pt}
                                                                                                                                      
\centerline{$^a$ {\sl Department of Physics, University of Cyprus,}}
\centerline{\sl CY-1678 Nicosia, Cyprus.}
\smallskip
\centerline{$^b$ {\sl Division of Science and Engineering, Frederick Institute of Technology,}}
\centerline{\sl CY-1303 Nicosia, Cyprus.} 

\vspace{24pt}
                                                                                                                                      
                                                                                                                                      
\centerline{{\bf Abstract}}
                                                                                                                                      
\noindent
{\narrower
We present results from numerical simulations of the Nambu$-$Jona-Lasinio model with 
an $SU(2) \otimes SU(2)$ chiral symmetry and $N_c=4,8$, and $16$ quark colors at nonzero temperature.
We performed the simulations by utilizing the hybrid Monte Carlo and hybrid Molecular Dynamics
algorithms. 
We show that the model undergoes
a second order phase transition. The critical exponents measured are consistent  
with the classical $3d$ $O(4)$ universality class and hence in accordance with the dimensional reduction 
scenario. We also show that the Ginzburg region is suppressed by a factor of $1/N_c$
in accordance with previous analytical predictions.
}

\bigskip
\noindent

\vfill
\newpage

\section{Introduction}
The behavior of symmetries at finite temperature is one of the most outstanding 
problems in major areas of particle physics such as cosmology, relativistic heavy-ion collisions,
and the quark-gluon plasma. In recent years, considerable work has been done on the physics of the
finite temperature chiral phase transition in QCD and related models. Since the problem of chiral 
symmetry breaking and its restoration is intrinsically non-perturbative, the number of techniques available
is limited.
Lattice simulations have so far provided us with most of our knowledge about this phenomenon. Various
lattice results indicate that the QCD transition is second order \cite{karsch}. However, due to problems with 
finite size effects near the critical temperature and the proper inclusion of fermions on the lattice, 
these calculations are still not fully satisfactory. Therefore, an important role is played by model 
field theories which incorporate certain basic features of QCD and they are easier to deal with. 
Such a theory is the Nambu$-$Jona-Lasinio (NJL) model. 
This  model was introduced in the sixties as a theory
of interacting nucleons \cite{nambu} and later it was reformulated in terms of quark degrees of freedom.
In this paper we present results from numerical simulations
of the NJL model near the thermal phase transition. 

A compelling idea based on 
universality and dimensional reduction \cite{pw}, is that in QCD with two massless
quarks the physics near the transition can be described by the three-dimensional $O(4)$-symmetric
sigma model. The reasoning behind the dimensional reduction scenario is based on the dominance
of the light degrees of freedom near the critical temperature $T_c$. In the imaginary time formalism 
of finite temperature field theory, a given system is defined on a manifold $S^1 \times R^d$, where $d$  is the number of spatial dimensions. 
The inverse temperature is the circumference of $S^1$. The boundary condition in the time direction 
is periodic for bosons and anti-periodic for fermions. At temperatures larger than the physical scales 
of the system, the nonzero Matsubara modes acquire a heavy mass. Therefore, the non-static configurations are strongly 
suppressed in the Boltzmann sum. If the original theory has fermions, 
all fermionic modes become massive due to the anti-periodicity in the temporal direction
and can be integrated over.
This is not to say that non-static modes have no effect. Rather, they generate counter-terms to the
three-dimensional theory. It is not yet clear whether the dimensional reduction is the correct 
description of the QCD phase transition.

In the continuum, the NJL model is described by the Euclidean Lagrangian density
\begin{equation}
{\cal L}  = \overline{\psi}_i \partial\hskip -.5em / \psi_i - \frac{g^2}{2N_c}
 \left[\left(\overline{\psi}_i\psi_i\right)^2-\left(\overline{\psi}_i
\gamma_5 \vec\tau\psi_i\right)^2\right],
\label{Lcont}
\end{equation}
where $\psi$ ($\bar{\psi}$) are the quark (anti-quark) fields 
and $\vec{\tau}\equiv(\tau_1,\tau_2,\tau_3)$ are the Pauli spin
matrices, which run over the internal $SU(2)_I$ isospin symmetry.
The index $i$ runs over $N_c$ quark colors and $g^2$ is the coupling
constant of the four-fermion interaction.
The model is chirally
symmetric under $SU(2)_L\otimes SU(2)_R$:
$\psi\to (P_LU+P_RV)\psi$,
where $U$ and $V$ are independent global $SU(2)$ rotations and the
operators
$P_{L,R}\equiv\frac{1}{2}(1\pm\gamma_5)$
project onto left
and right handed spinors, respectively. In other words the explicit flavor symmetry gives rise to 
$N_f=2$. It is also
invariant under $U(1)_V$
corresponding to a conserved baryon number.
The theory becomes easier to treat, both analytically and numerically, if we
introduce scalar and pseudo-scalar
auxiliary fields denoted by $\sigma$ and $\vec\pi$, respectively.
The bosonised Lagrangian is
\begin{eqnarray}
{\cal L}  =  \overline{\psi}_i\left(\partial\hskip -.5em / +\sigma+i\gamma_5
\vec\pi \cdot \vec\tau\right)\psi_i
+\frac{N_c}{2g^2}\left(\sigma^2+\vec\pi \cdot \vec\pi\right).
\end{eqnarray}
Since the theory has no gauge fields it fails to produce the physics of confinement.
The dynamic generation of fermion masses, though, brought about by the breaking of chiral symmetry
to $SU(2)_I$ at $g^2$ larger than a critical $g_c^2$ is one of the important properties of the model.
This implies the creation of both the Nambu-Goldstone bosons ($\pi$) and the chiral partner
sigma ($\sigma$), which are quark-antiquark bound states.
The interaction strength has a mass dimension -2, implying that the model is non-renormalizable. 
The triviality of the NJL model 
was also demostrated numerically in \cite{hands98}.
This model works well in the intermediate scale region as an effective theory of QCD with a momentum cutoff comparable
to the scale of chiral symmetry breaking $\Lambda_{\chi SB} \approx$ 1GeV.
Various authors used the NJL model to study properties of hadrons (for a review see 
\cite{hatsuda}).
It was also shown that the model has an interesting phase diagram in the
($T$, $\mu$) plane, where $\mu$ is the quark chemical potential. 
This phase diagram is in close agreement with predictions from other effective
field theories of QCD.
In the Hartree-Fock approximation, the theory has a tricritical point in the $(T,\mu)$ plane \cite{tricr}
when the quark current mass, the coupling $g$, and the momentum cut-off $\Lambda$
are chosen to have physically meaningful values.
Recently, lattice simulations showed that the ground-state at high $\mu$
and low $T$ is that of a traditional BCS superfluid \cite{hands02}.

A possible loophole of the dimensional reduction scenario is that in the NJL model (and in QCD) 
the meson states are composite quark$-$anti-quark bound states whose density and size increase as $T \rightarrow T_c$.
This may imply that the fermionic sub-structure is apparent on physical length scales and as a
consequence we may have a maximal violation of the bosonic character of mesons near $T_c$.
If this is indeed the case, then the quarks become essential degrees of freedom irrespective of how heavy
they are. Several authors observed \cite{rosen, kocic} that in the large $N_c$ limit, 
where quarks are the 
only degrees of freedom, the exponents 
of the finite temperature transition
are given by the Landau-Ginzburg Mean Field (MF) theory. 
The reasoning behind this result is that at finite temperature the zero fermionic modes are absent and therefore, 
contrary to a purely bosonic theory, the effect of making the temporal direction finite $(1/T)$ is to regulate the 
infrared behavior and suppress fluctuations.
However, inclusion of mesonic fluctuations in a non-perturbative
scheme that takes into account $1/N_c$ effects results in a first order finite temperature
phase transition \cite{oertel}. The first order transition discontinuity decreases with decreasing the 
cut-off that controls the strength 
of the mesonic fluctuations, but the nature of the transition remains unchanged even for very small values of the cut-off. 
It was also shown numerically in both the $d=(2+1)$ \cite{stephanov,cox} and $d=(3+1)$ \cite{chandra}
cases that the $Z_2$-symmetric model 
undergoes a second order finite temperature phase
transition. This transition belongs to the $(d-1)$-dimensional Ising model universality,
implying that the dimensional reduction
scenario is a valid description. 
In this paper, we study the case where the chiral symmetry is
continuum. 

In Sec.~2 we summarize the lattice formulation of the model. In Sec.~3
we present the observables  we used in the analysis of the critical properties of the theory. 
Finally, in Sec.~4 we present analysis of data in the vicinity of the critical temperature and 
in the low temperature broken phase. We show that the model undergoes a second order phase transition, which belongs to 
the $3d$ $O(4)$ universality class, and that the nontrivial scaling region is suppressed by a factor of $1/N_c$ 
in accordance with results presented in \cite{stephanov}.  

\section{Lattice Model} 
\label{sec:sec2}
The lattice action of the NJL model used in this study was the one first used in \cite{hands98},
\begin{eqnarray}
S & = & \sum_{\alpha=1}^N \sum_{xy}\left(\bar\chi_\alpha(x) M[\sigma,\vec\pi]_{xy}
\chi_\alpha(y)+\bar\zeta_\alpha(x)
M^\dagger[\sigma,\vec\pi]_{xy}\zeta_\alpha(y)\right) \nonumber \\
& + & {{2N}\over g^2}\sum_{\tilde x}(\sigma^2(\tilde x) +\vec\pi(\tilde x) \cdot\vec\pi(\tilde x)),
\label{eq:S}
\end{eqnarray}
where $\chi$, $\bar{\chi}$, $\zeta$, and $\bar{\zeta}$ are Grassmann-valued staggered fermion 
fields defined on the sites $x$ of a $(3+1)d$ Euclidean lattice. The scalar field $\sigma$ and the pseudoscalar
triplet $\vec{\pi}$ are defined on the dual lattice $\tilde{x}$ and the index $\alpha$ runs over $N$ staggered
fermion species. 
The kinetic operator $M$ is the usual one for four-fermion models with staggered fermions, modified 
to incorporate the $SU(2) \otimes SU(2)$ chiral symmetry and is given by 
\begin{eqnarray}
M_{xy} &=& \left({1\over2}\sum_\mu\eta_\mu(x)[\delta_{y,x+\hat\mu}
-\delta_{y,x-\hat\mu}]+m\delta_{xy}\right)\delta_{\alpha\beta}
\delta_{pq} \nonumber \\
&+& {1\over16}\delta_{xy}\delta_{\alpha\beta}\sum_{\langle\tilde x,x\rangle}
\Biggl(\sigma(\tilde x)+i\varepsilon(x)\vec\pi(\tilde
x) \cdot \vec\tau_{pq}\Biggr),
\label{eq:m}
\end{eqnarray}
where $m$ is the bare fermion mass and the $SU(2)_I$ indices $p,q$ are shown explicitly.
The phase factors $\epsilon(x) \equiv (-1)^{(x_0+x_1+x_2+x_3)}$ and $\eta_{\nu}=(-1)^{x_0+...+x_{\nu-1}}$
are the remnants of the $\gamma_5$ and $\gamma_{\nu}$ Dirac matrices on the lattice.
The symbol $\langle\tilde x,x\rangle$ denotes the set of $16$ dual sites
$\tilde x$ adjacent to the direct lattice site $x$. 

It can be easily shown that in the limit $m \rightarrow 0$, the action given by eq.~(\ref{eq:S}) has a global 
$SU(2)_L \otimes SU(2)_R$ \cite{hands98}. In the continuum limit, fermion doubling leads
to a physical content for the model of $8N$ fermion species - $4N$ described by $\chi$
and $4N$ by $\zeta$. 
The extra factor of $2$ over the usual relation for four-dimensional gauge theories 
arises from the impossibility of even/odd partitioning in the dual-site approach 
to lattice four-fermi models. 
We refer to these degrees of freedom as colors and hence define $N_c \equiv 8N$.
Further details about the lattice action are given in \cite{hands98}.

We performed  lattice simulations with $N_c=4,8$, and $16$. The $N_c=4$ simulation necessitates 
a fractional $N=0.5$. This can be achieved using the hybrid Molecular Dynamics (HMD) algorithm. 
More specifically, we used the ``R-algorithm'' of Gottlieb et al. \cite{R}, for which the systematic
error is $O(N_c^2 dt^2)$ \cite{hands98}, where $dt$ is the discrete simulation time-step. 
We checked that these systematic errors are smaller than 
the statistical errors of the various observables used to study the critical properties of the model.
On the other hand, the $N_c=8$ and $16$ simulations were performed by employing the exact hybrid Monte Carlo (HMC)
algorithm. The lengths of the Hamiltonian trajectories between momentum refreshments were 
drawn from a Poisson distribution with mean 2.0. For the inversion of matrix $M$ of eq.~(\ref{eq:m}),
we used the conjugate gradient method.

The simulations were performed on asymmetric lattices with $L_s$ lattice spacings $a$ in the three spatial dimensions, 
$L_t \ll L_s$ lattice spacings in the temporal direction, and volume $V \equiv L_s^3\cdot L_t$. The spatial 
extent must be much larger than the correlation length of the order parameter in order to obtain 
results that are close to the thermodynamic limit.
The temperature $T$ is given by $1/(L_ta)$. The temperature of the system may be altered
either by varying $\beta \equiv 1/g^2$, which amounts to varying the lattice spacing or by varying $L_t$.
In our simulations, we kept $L_t$ fixed and varied the lattice spacing.

\section{Observables}
In order to study the chiral symmetry of the model we work in the chiral limit. Hence, we choose 
not to introduce a bare quark mass into the lattice action. Without the benefit of this interaction,
the direction of symmetry breaking changes over the course of the simulation such that 
$\Sigma \equiv \frac{1}{V} \sum_x \sigma(x)$ and $\Pi_i \equiv \frac{1}{V} \sum_x \pi_i(x)$ 
average to zero over the ensemble. 
It is in this way that the absence of spontaneous symmetry breaking on a finite lattice is enforced. Another 
option is to introduce an effective order parameter $\Phi$ equal to the magnitude of the vector 
$\vec{\Phi}\equiv (\Sigma, \vec{\Pi})$. In the thermodynamic limit $\langle \Phi \rangle$ is equal to the true order parameter 
$\langle \sigma \rangle$ extrapolated to zero quark mass.

The Finite Size Scaling (FSS) method \cite{barber} is a well-established tool 
for studying phase transitions. We employ this method to study the 
critical behavior of the model on lattices available to us.
The correlation length $\xi$ on a finite lattice is limited by the size of the system and consequently
no true criticality can be observed.
The dependence of a given thermodynamic observable, $A$, on the
size $L_s$ of the box is singular. According to the FSS
hypothesis, in the large volume limit, $A$ is given by:
\begin{equation}\label{fssX}
A(t,L_s) = L_s^{\rho_A/\nu}Q_A(tL_s^{1/\nu}),
\end{equation}
where $t \equiv (\beta_c-\beta)/\beta_c$ is the reduced temperature,
$\nu$ is the exponent of the correlation length,
$Q_A$ is a scaling function that is not
singular at zero argument, and
$\rho_A$ is the critical exponent
for the quantity $A$.
Using eq.~(\ref{fssX}), one can determine such exponents 
by measuring $A$ for different values of $L_s$.

In the large $L_s$ limit, the FSS scaling form for the thermal average of the effective 
order parameter $\langle \Phi \rangle$ is given by
\begin{equation}
\langle \Phi \rangle =  L_s^{-\beta_{mag}/\nu}f_{\sigma}(tL_s^{1/\nu}).
\label{eq:magn1}
\end{equation}
Another quantity of interest is the susceptibility $\chi$ that is given, in the static limit of the
fluctuation-dissipation theorem, by
\begin{equation}
\chi = \lim_{L_s \to \infty} V[\langle \vec{\Phi}^2 \rangle - \langle \vec{\Phi} \rangle \cdot \langle \vec{\Phi} \rangle],
\end{equation}
where $V$ is the lattice volume.
For finite systems, the true order parameter $\langle \vec{\Phi} \rangle $ vanishes
and for $\beta \geq \beta_c$ the susceptibility is given by:
\begin{equation}
\chi = V\langle \Phi^2 \rangle. 
\label{eq:chi1}
\end{equation}
This relation should scale at criticality like $L_s^{\gamma/\nu}$. 
For small $L_s$, corrections to the FSS relations become important.

Furthermore, logarithmic derivatives of $\langle \Phi^j \rangle$ can give additional estimates for $\nu$.
It is easy to show that
\begin{equation}
D_j \equiv \frac{\partial}{\partial \beta} {\rm ln}\langle \Phi^j \rangle  =
\left[ \frac{\langle \Phi^j S_b \rangle}
{\langle \Phi^j \rangle} - \langle S_b \rangle \right],
\label{eq:derlog}
\end{equation}
where $S_b \equiv 2N\sum_{(\tilde x)}(\sigma^2(\tilde x) +\vec\pi(\tilde x) \cdot\vec\pi(\tilde x))$ is 
the purely bosonic part of the lattice action (eq.~(\ref{eq:S})). $D_j$  has a scaling relation
\begin{equation}
D_j = L_s^{1/\nu}f_{D_j}(tL_s^{1/\nu}).
\end{equation}

\section{Results}
\label{sec:sec3}
In this section we present the analysis of numerical results in both the vicinity of the 
finite temperature phase transition and the low temperature chirally broken phase. 
The HMD R algorithm was used to simulate the NJL model with 
$N_c=4$ whereas the HMC algorithm was 
used for $N_c=8$ and $16$.

\subsection{Finite Size Scaling}
\begin{figure}[htb]
\bigskip\bigskip \begin{center}
\epsfig{file=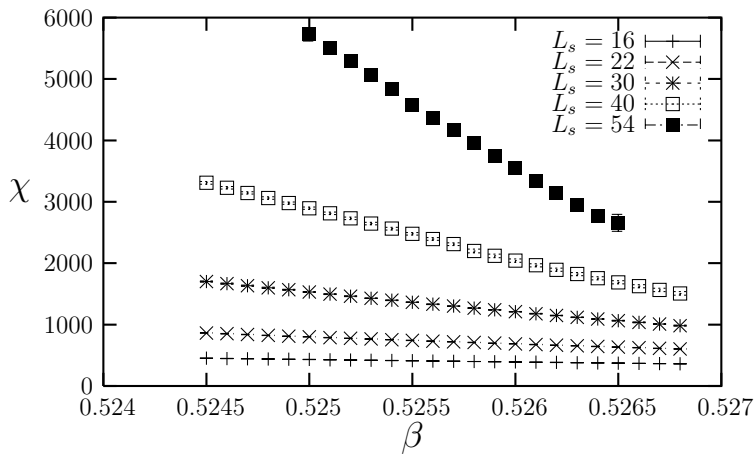, width=10cm}
\end{center}
\caption{The susceptibility $\chi$ vs. $\beta$ for different lattice volumes.}
\label{fig:chi1} 
\end{figure}

In this section we present the FSS analysis for the $N_c=8$ NJL model in the vicinity of the chiral phase transition. 
The lattice sizes ranged from $L_s=16$ to $L_s=54$ with fixed $L_t=4$. 
We used the histogram reweighting method \cite{histo} 
to perform our study most effectively. This 
enabled us to calculate the observables in a region of couplings around the simulation coupling.
We utilized the reweighting technique efficiently by performing simulations at 
slightly different couplings $\beta_i$ close to the critical coupling $\beta_c$. 
Details of the simulations are listed in Table~\ref{tab:t1}.
The expectation values of various observables $A_{L_s}^{(\beta_i)}(\beta)$
and the associated statistical errors $\Delta A_i$ were obtained by jackknife blocking.
They were then combined in a single
expression $A \equiv A_{L_s}(\beta)$ to obtain, as in \cite{janke}, 
\begin{equation}
A = \left[ \frac{A_1}{\Delta A_1} + ... + \frac{A_i}{\Delta A_i} \right](\Delta A)^2,
\end{equation}
where $\Delta A$ is given by 
\begin{equation}
\frac{1}{(\Delta A)^2} = \frac{1}{(\Delta A_1)^2} + ... + \frac{1}{(\Delta A_i)^2}.
\end{equation}
In Fig.~\ref{fig:chi1} we show the results of this procedure for the susceptibility $\chi$ defined in 
eq.~(\ref{eq:chi1}).
\begin{table}[]
\centering
\caption{Simulations for the FSS analysis.}
\medskip
\label{tab:t1}
\setlength{\tabcolsep}{1.0pc}
\begin{tabular}{ccc}
\hline \hline
$L_s $  &  $\beta_i$  &  Trajectories \\
\hline 
16  &  0.5250    &  220,000    \\
16  &  0.5260    &  440,000    \\
22  &  0.5250    &  200,000    \\
22  &  0.5260    &  290,000    \\
30  &  0.5250    &  100,000    \\
30  &  0.5260    &  160,000    \\
40  &  0.5250    &  65,000     \\
40  &  0.5260    &  66,000     \\
54  &  0.5255    &  40,000     \\
54  &  0.5260    &  50,000     \\
\hline \hline
\end{tabular}
\end{table}

\begin{figure}[b!]
\bigskip\bigskip \begin{center}
\epsfig{file=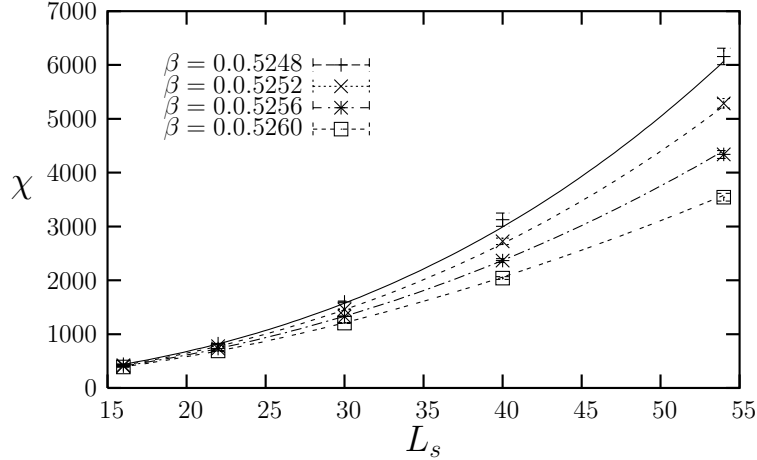, width=10cm}
\end{center}
\caption{The susceptibility $\chi$ vs. $L_s$ for different values of $\beta$.}
\label{fig:chi2} 
\end{figure}

Near $\beta_c$ the expansion of the FSS scaling relation for $\chi$, as in \cite{blote}, is:
\begin{eqnarray}
\chi & = & a(t) + b(x)L_s^{\gamma/\nu}, \nonumber \\
a(t) & \equiv &  a_0 + a_1t +a_2t^2 + ... \nonumber \\
b(x) & \equiv & b_0 + b_1x + b_2x^2 +... \;\;\; {\rm where}\;\;\; x \equiv tL_s^{1/\nu}.   
\label{eq:chi}
\end{eqnarray}
For large enough $L_s$ one can neglect $a(t)$. 
We fitted the data generated on lattices with $L_s=16,...,54$ to eq.~(\ref{eq:chi}) 
by including up to the linear terms in the expansions of $a(t)$ and $b(t)$. 
The data used in the fit are shown in Fig.~\ref{fig:chi1}.  
We extracted $\beta_c=0.52589(9)$, $\nu=0.747(20)$ and $\gamma/\nu=1.94(5)$ with $\chi^2/DOF=0.5$. 
The values of the exponents
are in good agreement with the $3d$ $O(4)$ universality exponents $\nu=0.7479(90)$ and $\gamma/\nu=1.97(1)$ 
reported in \cite{kanaya}. The possibility of a first order phase transition is clearly excluded, because
in such a scenario $\gamma/\nu=3$ (the number of the spatial dimensions). The high quality of this ``global''
fit is serious evidence
that the reported value for the critical coupling is accurate. 
If we neglect the $a(t)$ terms then in order to get $\beta_c=0.52572(16)$, $\nu=0.747(39)$ 
that are consistent with the results reported above 
we also have to discard the $L_s=16,22$ data. 
Inclusion of the smaller lattices gives results for $\nu$ and $\gamma/\nu$ that are significantly smaller than the 
expected $3d$ $O(4)$ exponents. Our analysis implies that finite size effects are relatively large 
and therefore one has to use
large lattices in order to reach the asymptotic FSS regime.
In Fig.~\ref{fig:chi2} we plot $\chi$ versus $L_s$ for four different values of $\beta$. 

\begin{table}[]
\centering
\caption{Results from fits to eq.~(\ref{eq:Dj}) for different ranges of $L_s$.}
\medskip
\label{tab:t2}
\setlength{\tabcolsep}{1.0pc}
\begin{tabular}{ccc}
\hline \hline
$D_j $  &    $\nu$  &  $\chi^2/{\rm DOF}$ \\
\hline
       & $L_s=22-54$ &           \\
\hline
$D_1$   &  0.750(8)  &  0.7   \\
$D_2$  &  0.749(6)   &   0.9  \\
$D_3$  &  0.747(6)   &  0.8  \\ 
\hline
        & $L_s=16-54$ &        \\
\hline
$D_1$   &  0.707(4)  & 1.6  \\
$D_2$   &  0.714(3)  & 1.7  \\
$D_3$  &   0.713(3)  & 1.6  \\
\hline \hline
\end{tabular}
\end{table}

\begin{table}[]
\centering
\caption{Results from fits to $D_j \sim L_s^{1/\nu}$ for $22 \leq L_s \leq 54$ at $\beta=0.52589$.}
\medskip
\label{tab:t3}
\setlength{\tabcolsep}{1.0pc}
\begin{tabular}{ccc}
\hline \hline
$D_j $  &    $\nu$  &  $\chi^2/{\rm DOF}$ \\
\hline \hline
$D_1$  &  0.721(23)   &  0.08   \\
$D_2$  &  0.721(21)   &  0.15  \\
$D_3$  &  0.722(20)   &  0.31  \\
\hline
\end{tabular}
\end{table}

\begin{figure}[t!]
\bigskip\bigskip \begin{center}
\epsfig{file=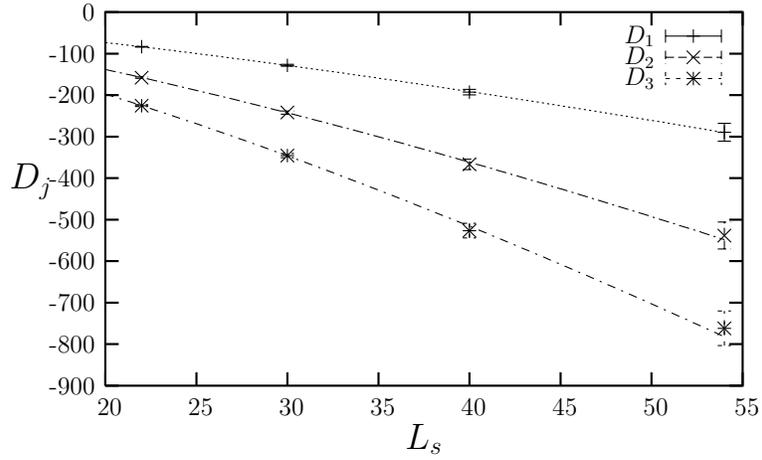, width=10cm}
\end{center}
\caption{$D_1,D_2$ and $D_3$ vs. $L_s$ at $\beta=0.52589$.}
\label{fig:dj} 
\end{figure}

Next, we present the FSS analysis of  the logarithmic derivatives $D_j$ defined in eq.~(\ref{eq:derlog}).
We fitted the $L_s \geq 22$ data to the linear expansion of the FSS relation 
of $D_j$, 
\begin{equation}
D_j(\beta,L_s) \simeq c_{1j}L_s^{1/\nu} + c_{2j}(\beta_c-\beta)L_s^{2/\nu},
\label{eq:Dj}
\end{equation}
with fixed $\beta_c=0.52589$ (obtained by the analysis of the susceptibility).
The results displayed in Table~\ref{tab:t2} show that the values of $\nu$ are in good
agreement with the $3d$ $O(4)$ result $\nu=0.7479(90)$ \cite{kanaya}. 
We should also note that if we include the $L_s=16$ data in the fits, the value of $\nu$
decreases and the quality of the new fits deteriorates as shown in Table~\ref{tab:t2}. 
The outcome of the fits to $D_j \sim L_s^{1/\nu}$ at $\beta=0.52589$
is shown in Table~\ref{tab:t3} and plotted in Fig~\ref{fig:dj}. As expected, these results are consistent 
with what we extracted from the global fits.  
The results of the analysis of the logarithmic derivatives confirm the conclusion of the
previous paragraph that large lattices are required in order to reach  
the asymptotic FSS regime. 
\begin{figure}[b!]
\bigskip\bigskip \begin{center}
\epsfig{file=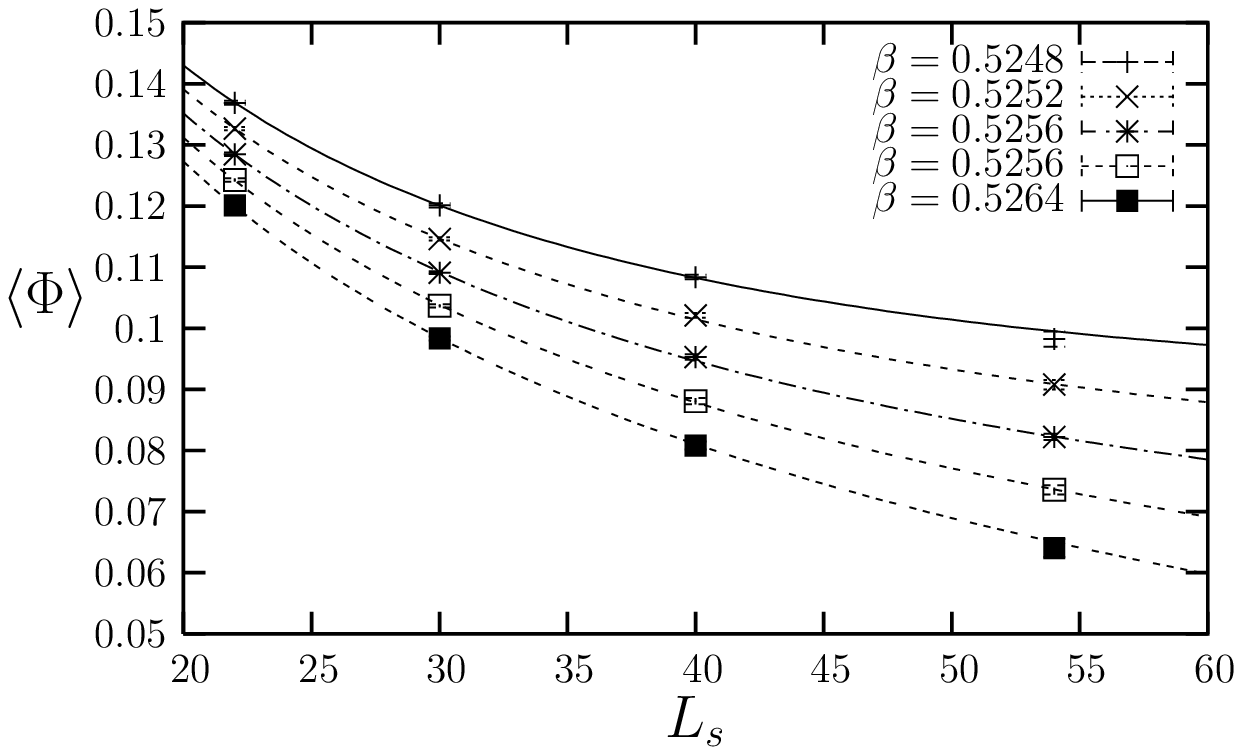, width=10cm}
\end{center}
\caption{Effective order parameter $\langle \Phi \rangle$ vs. $L_s$ at different values of $\beta$.}
\label{fig:fss.sigma}
\end{figure}

In this paragraph we present the FSS analysis of the effective order parameter $\langle \Phi \rangle$.
We fitted the $L_s \geq 22$ data
in the vicinity of the transition to a scaling function obtained
from the Taylor expansion of eq.~(\ref{eq:magn1}), up to a linear term,
\begin{equation}
\langle \Phi \rangle \simeq [c_1 + c_2(\beta_c-\beta)L_s^{1/\nu}]L_s^{\beta_{mag}/\nu}.
\end{equation}
We extracted $\nu=0.740(16)$, $\beta_{mag}/\nu=0.566(8)$, and $\beta_c=0.52593(4)$ with 
$\chi^2/DOF=0.43$. The measured value of $\nu$ is in good agreement with the $3d$ $O(4)$ result, 
whereas the value of $\beta_{mag}/\nu$ deviates from the $3d$ $O(4)$ result $\beta_{\rm mag}/\nu=0.5129(11)$ \cite{kanaya}.
This discrepancy in the value of $\beta_{\rm mag}$ might be due to higher order corrections to the FSS relation. 
In Fig.~\ref{fig:fss.sigma} we plot data used in this global fit versus the lattice spatial size $L_s$ at 
certain values of $\beta$. 
After including the quadratic term in the Taylor expansion of the FSS relation we got results consistent with those extracted 
from the linear expansion.  
Furthermore, attempts to add extra terms which could take into account corrections to FSS failed. The increase
in the number of fitting parameters resulted in bad signals for the various parameters.

\subsection{Broken Phase}

\label{sec:sec3.2}
\begin{figure}[t!]
\bigskip\bigskip \begin{center}
\epsfig{file=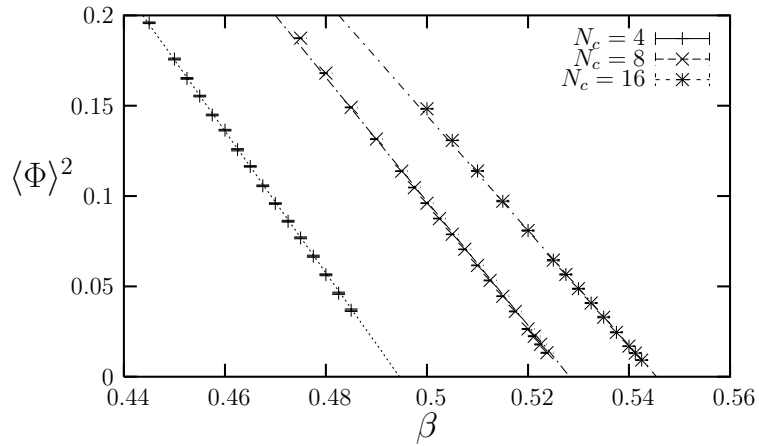, width=10cm}
\end{center}
\caption{$\langle \Phi \rangle^2$ vs. $\beta$ for $N_c=4,8$, and $16$.
The straight lines represent fits to the data in the mean field regions of the different $N_c$'s.}
\label{fig:sigma2}
\end{figure}
                                                                                                                       
\label{sec:sec3.2}
\begin{figure}[b!]
\bigskip\bigskip \begin{center}
\epsfig{file=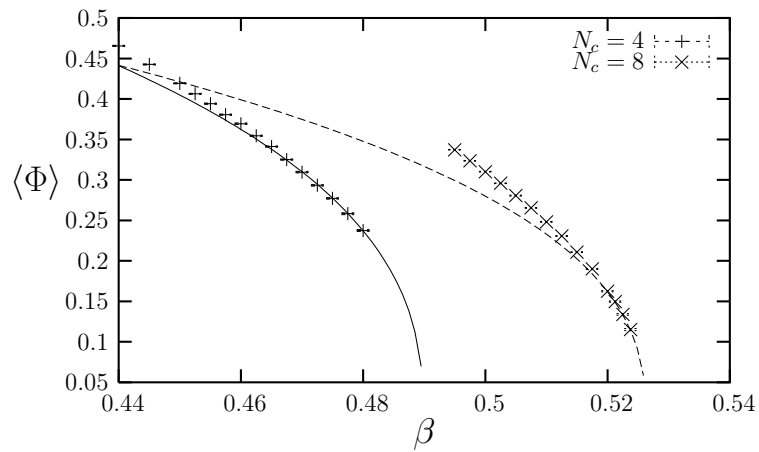, width=10cm}
\end{center}
\caption{$\langle \Phi \rangle$ vs. $\beta$. The continuum and dashed
lines represent fits to the data in the $3d$ $O(4)$ scaling regions of $N_c=4,8$ repsectively.}
\label{fig:o4.scaling}
\end{figure}
We performed simulations in the broken phase
in order to study the dependence of $\langle \Phi \rangle$ on $\beta$ at three different values of $N_c=4,8$, and 
$16$.
According to analytical and numerical evidence \cite{stephanov},
the non-trivial scaling region of the finite temperature transition in 
the three-dimensional Gross-Neveu model
is suppressed by a factor of $1/\sqrt{N_c}$.  
The reasoning behind this result is that for finite $N_c$, close enough to $T_c$  where 
the bosonic fluctuations become important, MF theory 
breaks down due to self-inconsistency. Hence, although the fermions do not 
affect the universal critical properties of the theory, they do influence
the width of the actual critical region. 
It was also shown in \cite{stephanov}
that in four-dimensions the non-trivial scaling region is suppressed
by a factor of $1/[N_c \ln(L_t)]$. The logarithmic dependence on $L_t$ 
is an outcome of the triviality of the four-dimensional theory.

We performed simulations on lattices with fixed $L_t=4$ and $L_s=54$. 
The Landau-Ginsburg MF theory predicts $\beta_{\rm mag} =1/2$. Therefore, in the 
MF regions of different $N_s$'s the data for $\langle \Phi \rangle^2$ are expected to fit 
well to straight lines: $\langle \Phi \rangle^2 \sim ({\rm const.} - \beta)^1$. 
The results for $\langle \Phi \rangle^2$ versus $\beta$  are presented in Fig.~\ref{fig:sigma2}.
The $N_c=16$ data shown in Fig.~\ref{fig:sigma2}
fit well to the MF scaling for all the values of $\beta$. 
In the $N_c=4$ and $8$ cases however, for sufficiently small $\langle \Phi \rangle$,  there is a crossover 
from MF scaling into the $3d$ $O(4)$ universality. This is demonstrated more clearly in Fig.~\ref{fig:o4.scaling} 
where it is shown that the data for small values of $\langle \Phi \rangle$, fit well to
\begin{equation}
\langle \Phi \rangle = a(\beta_c-\beta)^{\beta_{\rm mag}},
\label{eq:scaling}
\end{equation}
where $\beta_{\rm mag}$ is fixed to its $3d$ $O(4)$ value of 0.3836 \cite{kanaya}.
The continuum and dashed lines in Fig.~\ref{fig:o4.scaling}
represent fits to the data in the non-trivial scaling regions for $N_c=4$ and $N_c=8$, respectively.  
For $N_c=4$, we extracted $\beta_c=0.4901(3)$ by fitting the data in the range $0.470 \leq \beta \leq 0.480$.
For $N_c=8$, we extracted $\beta_c=0.5263(1)$ by fitting the data for couplings $0.52000 \leq \beta \leq 0.52375$.
In Table 3 we show the $\chi^2/DOF$ for different sets of data fitted to eq.~(\ref{eq:scaling}). 
It is clear that 
the fit quality deteriorates as we add data points at lower temperatures. 
Hence, we conclude that for $N_c=4$ the crossover out of the 
$3d$ $O(4)$ scaling region occurs roughly at $\beta_{\rm cross}=0.465$ 
with $\langle \Phi \rangle_{\rm cross}=0.34 $ 
and for $N_c=8$
at $\beta_{\rm cross}=0.5175$ with $\langle \Phi \rangle_{\rm cross}=0.190$. At this point we should note
that in the MF region the order parameter is equal to the thermal fermion mass \cite{rosen}. Therefore, it acts 
as an inverse correlation length. Given that our 
definition of the crossover from one scaling region 
to another is a bit loose, we believe that our results are consistent with the analytical prediction 
that the scaling region is suppressed by a factor of $1/N_c$ \cite{stephanov}.

We also fitted the $N_c=8$ effective order parameter data for $\beta=$0.5200, 0.52125, and 0.5225 to 
eq.~(\ref{eq:scaling}) with the critical coupling fixed 
to $\beta_c=0.52589$ (extracted from the FSS analysis of the susceptibility). 
We found $\beta_{\rm mag}$ to be 0.350(15), which is 
roughly two standard deviations below the $3d$ $O(4)$ result $\beta_{\rm mag} = 0.3836(46)$. 
We believe that this small discrepancy 
is due to finite size effects which cause an increase in the value of $\langle \Phi \rangle$. 
Given that $\beta_c$ is fixed, the fitting fuction is forced to become more abrupt, resulting in 
a smaller value for $\beta_{\rm mag}$.

\begin{table}[]
\centering
\caption{Simulations for the FSS analysis.}
\medskip
\label{tab:t4}
\setlength{\tabcolsep}{1.0pc}
\begin{tabular}{ccc}
\hline \hline
No. of points    &    $\beta$ range  &  $\chi^2/{\rm DOF}$ \\ 
\hline 
& $N_c=4$ &  \\
\hline
4   &  0.4725-0.4800  &  0.10  \\
5   &  0.4700-0.4800  & 0.17  \\
6   &  0.4675-0.4800  & 0.58  \\
7  &  0.4650-0.4800   & 2.50   \\
8  &  0.4625-0.4800   & 3.2    \\
\hline
& $N_c=8$ &  \\
\hline
3   &  0.52125-0.52375 & 0.08  \\
4   &  0.52000-0.52375 & 0.07  \\
5   &  0.5175-0.52375  & 7.3   \\
\hline \hline
\end{tabular}
\end{table}

\section{Conclusions}
We showed with relatively good precision that the NJL model with four lattice spacings in the 
temporal direction undergoes a second order phase transition. 
The critical indices were extracted by employing FSS analysis in the vicinity of $T_c$ and analysis of data
generated on large lattices in the low temperature chirally broken phase. Our results are consistent 
with the $3d$ $O(4)$ classical spin model exponents
and hence in accordance with the dimensional reduction scenario \cite{pw}. 
We summarize our results in Table~\ref{tab:t5} 
and compare them with those of the classical $3d$ $O(4)$ Heisenberg spin model \cite{kanaya}. The values 
and the error bars of the exponents presented in Table~\ref{tab:t5} take into account slightly 
different predictions for each exponent measured by different methods in this study. 

We also verified numerically a previous analytical result \cite{stephanov} 
that the actual critical region is suppressed by a factor of $1/N_c$. Outside this region,
where fluctuations are negligible, we have Landau-Ginsburg MF 
scaling  predicted by the large-$N_c$ calculations. This result should be contrasted with 
recent numerical results in strong coupling QCD, where the width of the actual critical region  
does not depend on $N_c$ \cite{strongQCD}. 

Furthermore, it is useful to comment on the phenomenological significance of the parameters used
in this simulation. Within the framework of the Hartree approximation (for details see \cite{hands02}),
it can be easily shown 
that if one fixes $F_{\pi}$ to its experimental value of 93MeV and the constituent quark mass
to a physically reasonable value of 400MeV then the inverse lattice spacing $a^{-1}$ at the critical
coupling $\beta_c=0.5259$ (for $L_t=4$) of the $N_c=8$ nonzero temperature phase transition is of the order of 1GeV. 

\begin{table}[]
\centering
\caption{Critical exponents of the $3d$ $O(4)$ Heisenberg model obtained in \cite{kanaya}
and critical exponents obtained from this study.}
\medskip
\label{tab:t5}
\setlength{\tabcolsep}{1.0pc}
\begin{tabular}{ccc}
\hline \hline
                & $3d$ $O(4)$ spin model exponents   &  This study \\
\hline
$\nu$             &  0.7479(90)      &    0.75(4)   \\
$\gamma$          &  1.477(18)       &    1.46(6)   \\
$\beta_{\rm mag}$ &  0.3836(46)      &    0.38(4)   \\
\hline \hline
\end{tabular}
\end{table}

\section*{Acknowledgements}
Discussions with Shailesh Chandrasekharan, Simon Hands, John Kogut, and Misha Stephanov are greatly appreciated.

\end{document}